\documentclass[universe,review,accept,moreauthors,pdftex]{Definitions/mdpi} 

\usepackage{color}

\firstpage{1} 
\pubvolume{1}
\issuenum{1}
\articlenumber{0}
\pubyear{2021}
\copyrightyear{2020}
\datereceived{} 
\dateaccepted{} 
\datepublished{} 
\hreflink{https://doi.org/} 

\Title{Astroparticle physics with compact objects}

\TitleCitation{Astroparticle physics with compact objects}

\def\ghost#1{\vrule height#1 depth#1 width0pt \displaystyle}

\newcommand{\gag}{g_{a\gamma\gamma}}

\Author{Peter Tinyakov $^{1}$\orcidA{}, Maxim Pshirkov $^{2,3,4}$\orcidB{} and Sergei Popov $^{2}$*\orcidC{}}

\AuthorNames{Peter Tinyakov, Maxim Pshirkov and Sergei Popov}

\AuthorCitation{Tinyakov, P.; Pshirkov, M.; Popov, S.}

\address{%
$^{1}$ \quad Universit\'e Libre de Bruxelles, bvd. du Triomphe, CP225, 1050 Brussels, Belgium\\
$^{2}$ \quad Sternberg Astronomical Institute, Lomonosov Moscow State University, 119234 Moscow, Russia\\
$^{3}$ \quad Institute for Nuclear Research of Russian Academy of Sciences, 60th October Anniversary Prospect, 7a, Moscow, 117312, Russia\\
$^{4}$ \quad P. N. Lebedev Physical Institute of the Russian Academy of Sciences, Pushchino Radio Astronomy Observatory, PRAO, Pushchino, 142290, Russia}

\corres{Correspondence: sergepolar@gmail.com; Tel.:  +7-495-939-5006 (S.P.)}

\abstract{Probing the existence of hypothetical particles beyond the Standard model often deals with extreme parameters: large energies, tiny cross-sections, large time scales, etc. Sometimes laboratory experiments can test required regions of parameter space, but more often natural limitations leads to poorly restrictive upper limits. In such cases astrophysical studies can help to expand the range of values significantly. Among astronomical sources, used in interests of fundamental physics, compact objects --- neutron stars and white dwarfs, --- play a leading role. We review several aspects of astroparticle physics studies related to observations and properties of these celestial bodies. Dark matter particles can be collected inside compact objects resulting in additional heating or collapse. We summarize regimes and rates of particle capturing as well as possible astrophysical consequences. Then we focus on a particular type of hypothetical particles --- axions. Their existence can be uncovered due to observations of emission originated due to Primakoff process in magnetospheres of neutron stars or white dwarfs. Alternatively, they can contribute to cooling of these compact objects. We present results in these areas, including upper limits based on recent observations. }


\begin{document}

\section{Introduction}

It is a commonplace to state that compact objects --- white dwarfs, neutron stars, and black holes, --- are unique natural laboratories which allows us to study matter and various processes under extreme conditions. However, this is indeed true and quite often best limits on different hypothesis in fundamental physics beyond standard scenarios are given by observations of these astrophysical sources. 

Leaving black holes (BHs) aside, 
neutron stars (NSs) and white dwarfs (WDs) with their high density and temperature of interiors and large external and internal magnetic fields in many respects are well-suited for studies of  interactions with even very elusive particles (like dark matter candidates, axions and axion-like particles, etc.) and for their production.

Internal densities of NSs can reach up to $\lesssim 10$ nuclear saturation density \cite{2007ASSL..326.....H}. This allows to obtain a reasonable rate of interaction even with particles with very small cross-section reducing their energy in the process and eventually trapping them inside compact objects. Despite WDs have about seven orders of magnitude lower density, they also can capture weakly interacting particles. 

High densities combined with high temperatures result in non-trivial cooling processes of compact objects which are born very hot. 
The main cooling channel for many of them is related to neutrino emission \cite{2004ARA&A..42..169Y}. However, it was also proposed that axions can play a role in thermal energy losses in both NSs and WDs (see , e.g. \cite{1992ApJ...392L..23I, 2016PhRvD..93f5044S}).

In the case of magnetars \cite{2015RPPh...78k6901T} and magnetic WDs (MWDs) \cite{2015SSRv..191..111F} magnetic moments  can be as high as $10^{33}$~G~cm$^3$. Magnetospheres of such compact objects can interact with axions via the Primakoff process, thus working as a kind of a haloscope \cite{1996slfp.book.....R}. Typically, magnetic fields of other astronomical objects (or in realistic medium) are not that high to produce significant effect for discussed parameters of particles. Thus, NSs and WDs can give us a unique opportunity to discover axions with purely astrophysical methods.  

In this review BHs generally are not discussed, together with different exotic solutions like boson stars \cite{1986PhRvL..57.2485C}, fermion stars \cite{1987PhRvD..35.3678L}, pion stars \cite{2018PhRvD..98i4510B}, etc. Instead we focus on three abovementioned aspects linking particle physics beyond the Standard model with properties of NSs and WDs. 

At first, in Sec. 2, we discuss how dark matter (DM) particles can appear inside NSs and WDs due to capturing by progenitors or already after the compact object formation. Then, studies of the main possible consequences --- heating and collapse, --- of particles capture are reviewed. 

In the next section we turn  to a   specific type of hypothetical particles --- axions (including so-called axion-like particles, ALPs). We present theory of axion interaction with magnetospheres of NSs.  We discuss  interactions with axions of cosmological origin that could constitute dark matter and those produced in NSs. Interaction of axion clusters with magnetospheres is also discussed in the context of models of fast radio bursts (FRBs) origin.

Finally, in Sec. 4 we come to WDs and review possible role of axions in cooling of these compact objects after which present several upper limits on properties of these particles based on observations of WDs.


\newcommand{\pt}[1]{\textcolor{blue}{[PT: #1]}}

\section{Compact objects and dark matter}
\label{sec:compact-objects-dark}

According to the theory of structure formation, at galactic and larger scales
the distributions of baryons and dark matter are correlated. Stars are
therefore formed in regions of DM overdensities. As any gravitating objects,
they have some amount of DM gravitationally bound to them. They may also
accumulate DM during their lifetime. While the total amount of the accumulated
DM stays tiny, it may nevertheless have observable effects. In this section we
review the DM capture mechanisms and estimate the amount of DM that may be
accumulated by a star in different conditions. We then discuss possible
observational signatures that may result.

\subsection{Capture of DM by stars}
\label{sec:capture-dm-compact}

Stars capture DM at formation and during their lifetime by two different
mechanisms. The second one --- accumulation during the star lifetime --- requires
non-zero DM-to-nucleon interactions for capture; the first mechanism is gravitational at the stage of capture, but still requires DM interactions with 
nucleons for the thermalization of the captured DM. The two mechanisms generally give comparable contributions to the total amount of DM accumulated by a star.  We start with
the second one as it is more extensively studied in the literature and allows
for simple analytical estimates.

\subsubsection{Capture during star lifetime}
\label{sec:capture-lifetime}

To gain a qualitative understanding of the capture mechanism, consider an isolated star\footnote{An enhancement of capture rate may be expected in binary systems, but only by a factor of a few \cite{Brayeur:2011yw}.} embedded in a gas of non-interacting DM
particles, as would be the case for stars in galactic halos. Assume the DM
distribution in velocities is Maxwellian, characterised by a velocity
dispersion $\bar v$. Assume as well that the star is at rest with respect to
the DM distribution. This is {\em not} the case for most of the stars in
halos as their typical velocities are the same as those of DM particles, that
is, of order $\bar v$.  Also, the DM distribution in velocities may not be
Gaussian in a realistic halo. We will discuss later the effect of  these
complications on the capture rate.

The star forms a gravitational potential well where DM particles are confined
provided they have negative total energy. A particle from the DM halo that
originally has a positive energy can lose it in collisions with the stellar
material and become gravitationally bound. Since the collisions happen while a
particle is inside the star, it settles on a star-crossing orbit and will
eventually come back and cross the star again. The energy losses will thus
continue until the particle reaches an equilibrium with the matter inside the
star. The rate of capture is  determined by the number of particles that
get gravitationally bound after the first star crossing. 

It follows immediately from this picture that there is an upper bound
on the capture rate $F$ independent of the energy loss mechanism. It is given by
the rate of star crossings by the DM particles. For a given particle with 
asymptotic velocity, $v$, the cross section of the star crossing is 
\begin{equation}
\sigma_{\rm cross} = \pi R_*^2 \left( 1 + {2GM_* \over R_*v^2} \right) = 
\pi R_*^2 \left( 1 + {v_{\rm esc}^2\over v^2} \right),
\label{eq:sigma_crossing}
\end{equation}
where $R_*$ and $M_*$ are the star radius and mass, $G$ is the gravitational
constant, and $v_{\rm esc}$ the escape velocity from the surface of the
star. Here for simplicity the Newtonian gravity has been assumed, which gives a reasonable
approximation even for a NS. The account for General Relativity corrections changes the capture rate by a factor of order 1 
\cite{Kouvaris:2007ay}. The second term in the parenthesis is
the result of gravitational focusing; this term is dominant in most
cases. Averaging this expression over the Maxwellian distribution of the
velocity dispersion $\bar v$ yields the maximum value of the DM capture rate $F$, 
\begin{equation}
F_{\rm max} = \left({8\pi\over 3}\right)^{1/2} 
n R_*^2 \bar v 
\left( 1+ {3 v_{\rm esc}^2\over 2 {\bar v}^2}\right)
\simeq \sqrt{6\pi} n R_*^2 v_{\rm esc}^2/{\bar v} = 
\sqrt{6\pi} n {R_* R_g\over {\bar v}}c^2, 
\label{eq:max_rate}
\end{equation}
where $n$ is the DM number density and 
$R_g=2GM_*/c^2$ 
is a stellar gravitational
radius. In the final expression we have neglected the first term in the
r.h.s. of eq.~(\ref{eq:sigma_crossing}) as compared to the second
one. Detailed calculations can be found in Refs.~\cite{Press:1985ug,Gould:1987ju,Gould:1987ww}. Eq.~(\ref{eq:max_rate})
suggests that loosely bound haloes with smaller velocity dispersion
give higher capture rates.

To get an idea of how much DM can {\em possibly} be captured by this mechanism
let us substitute the numbers. For a Sun-like star in the typical Galactic environment,
assuming the lifetime of 10~Gyr eq.~(\ref{eq:max_rate}) gives for the total
mass of the accumulated DM $M_{\rm tot}\sim 5\times 10^{47}~{\rm GeV} \sim
5\times 10^{-10}~M_\odot$. For a NS in the same conditions $M_{\rm
  tot}\sim 10^{43}~{\rm GeV} \sim 10^{-14}~M_\odot$. The upper limits may be higher by 2-3 orders of magnitude in an environment with higher DM density and smaller velocity. 
 
The capture rate, however, is usually further suppressed for two reasons: not
every particle that crosses the star may get scattered, and even if it does,
not every particle loses enough energy to become gravitationally bound. The
scattering is controlled by the cross section of 
DM interactions with nucleons.  There is a
critical value of this cross section $\sigma_{\rm cr}$ at which the star
becomes opaque to DM particles, that is, at which a DM particle scatters once
per crossing on average.  It is estimated as $\sigma_{\rm cr} = R_*^2 m_p/M_*$
and has the following value for the Sun, a WD and a NS,
respectively:
\begin{eqnarray}
\nonumber
\sigma_{\rm cr,~Sun} &&\simeq 4\times 10^{-36} {\rm cm}^2,\\
\nonumber
\sigma_{\rm cr,~WD} &&\simeq 4\times 10^{-40} {\rm cm}^2,\\  
\label{eq:sigma_cr}
\sigma_{\rm cr,~NS} &&\simeq 8\times 10^{-46} {\rm cm}^2.
\end{eqnarray}
For NSs and WDs we have neglected the effects of matter degeneracy; they will be
commented on later.  If the DM-nucleon cross section is smaller than the
critical one, the capture rate gets proportionally suppressed. Note that this
is a likely case as the experimental bounds on the DM-nucleon cross section
$\sigma_{\rm DM-n}$ are becoming quite strict. In the case of 
the spin-independent interactions they are at the level of $10^{-46}$~cm$^2$
at DM mass around $50$~GeV \cite{XENON:2018voc},
while for the spin-dependent interactions the limits on the cross section 
are at the level of $10^{-40}-10^{-41}$~cm$^2$ \cite{XENON:2019rxp,LUX:2017ree,PandaX-II:2018woa,PICO:2019vsc}. The constraints become much weaker for DM masses below $10$~GeV and higher than several TeV. 

Consider now the effect of energy losses on the capture rate. A particle
crossing the star gets trapped if it loses the energy $E_{\rm loss}$ which is larger
than its asymptotic energy. If $E_{\rm loss} \gtrsim m \bar v^2/2$ most of the
particles crossing the star get trapped and no additional suppression results
as compared to eq.~(\ref{eq:max_rate}). In the opposite case
only a fraction of particles that have sufficiently small asymptotic velocity
$v<\bar v$ such that $mv^2/2<E_{\rm loss}$ will become gravitationally
bound. Given the distribution of DM particles over velocities it is a matter
of a direct calculation to find the capture as a function of $E_{\rm
  loss}$. Assuming all particles lose the same energy one gets the capture rate  
\begin{equation}
F = \sqrt{6\pi}  {\rho_{\rm DM} R_* R_g\over m {\bar v}(1-R_g/R_*)} 
\left[ 1 - \exp \left(- {3 E_{\rm loss} \over m \bar v^2}\right)\right]f. 
\label{eq:capture}
\end{equation}
Here 
$\rho_{\rm DM}$ is
the DM ambient density  and $f$ is the
fraction of particles that scatter among all particles
crossing the star
\begin{equation}
\begin{array}{ll}
f=\sigma/\sigma_{\rm cr} & \text{if $\sigma < \sigma_{\rm cr}$},\\
f= 1 & \text{otherwise.}
\end{array}
\label{eq:f}
\end{equation}
For completeness, we have also included in eq.~(\ref{eq:capture}) the relativistic correction $1/(1-R_g/R_*)$ \cite{Kouvaris:2007ay} which however is only noticeable in case of NSs where it leads to a slight enhancement of the rate. We will disregard it in what follows. 

The dependence of the capture rate on the energy loss $E_{\rm loss}$ is
controlled by the factor in the square brackets. When the energy loss in a
single star crossing is of order or larger than the typical asymptotic energy
$m \bar v^2/2$, this factor is close to 1 and gives no extra suppression of
the capture rate. In this case, and assuming all particles scatter ($f=1$) one
recovers the maximum rate~(\ref{eq:max_rate}). Note that in this regime the
{\em mass} capture rate $mF$ is independent of the DM mass $m$ at a given
ambient DM density $\rho_{\rm DM}$. In the opposite limit when the energy
losses are small, $E_{\rm loss}\ll m \bar v^2$, the capture rate is further
suppressed.  In this regime the mass capture rate is inversely proportional to
the DM mass. Note also that in this case the dependence on the DM asymptotic
velocity is much stronger, $F\propto 1/\bar v^3$. In summary,
\begin{equation}
F \simeq \left\{
\begin{array}{ll}
\ghost{18pt}
\sqrt{6\pi}  {\rho_{\rm DM} R_* R_g\over m {\bar v}} 
\,f & \quad \text{at large $E_{\rm loss}\gg m\bar v^2$}, \\
\ghost{18pt}
\sqrt{6\pi}  {\rho_{\rm DM} R_* R_g\over m {\bar v}} 
{3 E_{\rm loss} \over m \bar v^2}\,f & \quad \text{at small
 $E_{\rm loss}\ll m\bar v^2$} .
\end{array}\right.
\label{eq:rate-final}
\end{equation}

The extra suppression factor $\propto E_{\rm loss}/(m \bar v^2)$ at small 
 $E_{\rm loss}$ arises
because in this case only a small fraction of the DM particles ---
those that have $m v^2 \lesssim E_{\rm loss}$ --- can lose enough energy in
one star crossing and become gravitationally bound. Clearly, in this regime
only the low-velocity part of the DM distribution is important. The
velocity dependence of the capture rate (\ref{eq:rate-final}) originates in
this case from the volume element in the velocity space $d^3v$ and will be the same for
any velocity distribution that, like the Maxwellian one, goes to a constant at
small $v$. Therefore, regardless of the details of the DM velocity
distribution, the capture rate is proportional to $F\propto 1/\bar v$ and is
given by eq.~(\ref{eq:max_rate}) times the probability of scattering $f$ for
small $\bar v$, and changes to $F\propto 1/\bar v^3$ as given by the small 
 $E_{\rm loss}$ case of eq.~(\ref{eq:rate-final}) 
with possibly a different numerical coefficient. 

When expressed in terms of the energy loss $E_{\rm loss}$ and the probability
of scattering $f$, eqs. (\ref{eq:rate-final})  are
completely general and apply to any type of DM and/or stars. The specific
features of the models determine these two quantities. The probability $f$ is 
determined by the scattering cross section through eq.~(\ref{eq:f}) for
particle DM, for primordial black holes $f=1$ .

Consider now the energy losses $E_{\rm loss}$. We start with the simplest case
of particle DM crossing an ordinary star. A typical kinetic energy of a DM
particle away from the star is $m \bar v^2/2$. When the particle falls onto
the star it accelerates and picks the velocity of order of the escape
velocity. For a Sun-like star the latter is $\sim 600$~km/s, which is about
factor 3 higher than the dispersion velocity in our Galaxy at the position of
the Sun. We can therefore estimate the particle velocity while crossing the star as
the escape velocity $v_{\rm esc} \propto \sqrt{R_g/R_*}$. At the temperature of
order $10^7$~K the protons in the core of a Sun-like star move with the
thermal velocity $\sqrt{3kT/m_p} \sim 500$~km/s, and with smaller velocities
away from the core. For the sake of the estimate we may, therefore, consider
the collision of DM particles with protons at rest. Assume for simplicity that
DM particles are much heavier than protons. The typical energy transfer in the
collision, in the frame of the star, is then $E_{\rm loss}\sim m_pv_{\rm
  rel}^2/2$, where $v_{\rm rel}\sim v_{\rm esc}$ is the relative velocity of
particles. Depending on the environment this energy loss may be larger or
smaller than the particle asymptotic energy. In our Galaxy and 
for DM particles heavier than $\sim 10$~GeV the energy loss is
smaller than typical asymptotic energy, so we are in the suppressed small
$E_{\rm loss}$ regime 
of eq.~(\ref{eq:rate-final}). In dwarf galaxies with typical
velocity dispersion $\sim 10$~km/s this regime occurs for DM masses $\gtrsim
500$~GeV, otherwise the large $E_{\rm loss}$ regime of 
eq.~(\ref{eq:rate-final}) gives the right estimate of the rate. Thus, in these
two cases we have for the mass capture rate:
\begin{eqnarray}
\nonumber
&& \text{Sun-like, Milky Way:} \\
\label{eq:F_Sun_MW}
&&\qquad \ghost{15pt} mF \sim 2\times 10^{30}\, {\rm GeV/s} 
\left({\rho_{\rm DM}\over 0.5\,{\rm GeV/cm}^3 }\right) 
\left({10\,{\rm GeV}\over m} \right)\,f, 
\\
\nonumber
&&\text{Sun-like, dwarf galaxies, $m \gtrsim 500$~GeV:}  \\
\label{eq:F_Sun_dwarf_hm}
&&\qquad \ghost{15pt} mF \sim   
9\times 10^{34}\, {\rm GeV/s} 
\left({\rho_{\rm DM}\over 100\, {\rm GeV/cm}^3 }\right) 
\left({500\,{\rm GeV}\over m} \right)\,f,
\\
\nonumber
&&\text{Sun-like, dwarf galaxies, $m \lesssim 500$~GeV:}  \\
\label{eq:F_Sun_dwarf_lm}
&&\qquad \ghost{15pt} mF \sim   
9\times 10^{34} \, {\rm GeV/s} 
\left({\rho_{\rm DM}\over 100\, {\rm GeV/cm}^3 }\right) 
\,f,
\end{eqnarray}
The numerical values in eqs.~(\ref{eq:F_Sun_MW})--(\ref{eq:F_Sun_dwarf_lm}) 
are close to the maximum achievable capture rates for the Sun-like star in the 
corresponding environment. Note, however, that in view of
eqs.~(\ref{eq:sigma_cr}), (\ref{eq:f}) for the Sun-like star the scattering
probability is $f\lesssim 10^{-10}$ if current experimental bounds on the spin-independent DM-nucleon cross section are assumed. 

In the case of compact objects --- WDs or NSs, --- the escape velocities are substantially
higher, $v_{\rm esc} \sim 6\times 10^3$~km/s and $v_{\rm esc} \sim 0.5 c$,
respectively. In this case we are always in the large $E_{\rm loss}$ regime of
eq.~(\ref{eq:rate-final}) for DM masses below $\sim 1$~TeV. The capture rate
is then estimated as follows, 
\begin{eqnarray}
&& \text{WD:} 
\label{eq:F_WD}
\quad \ghost{15pt} mF \sim 2\times 10^{28}\,{\rm GeV/s} 
\left({\rho_{\rm DM}\over 0.5\, {\rm GeV/cm}^3 }\right) 
\left({220\, {\rm km/s}\over \bar v} \right)\,f, 
\\
&&\text{NS:}  
\label{eq:F_NS}
\quad \ghost{15pt} mF \sim   
2.5\times 10^{26}\, {\rm GeV/s} 
\left({\rho_{\rm DM}\over 0.5\, {\rm GeV/cm}^3 }\right) 
\left({220\, {\rm km/s}\over \bar v} \right)\,f.
\end{eqnarray}
In the case of compact stars the critical cross sections are smaller,
cf. eq.~(\ref{eq:sigma_cr}), so the scattering probability $f$ is less
suppressed than in the case of Sun-like stars. For NSs, assuming
current experimental limits on the scattering cross section, it may reach
values close to 1 even for the spin-independent cross section. In all cases, however, 
the accumulation rates are small: only a tiny fraction of $M_\odot$ can be
accumulated over the age of the Universe. 

A number of subtleties arises in the case of capture by NSs because NS matter is degenerate, and because the scattering happens at semi-relativistic velocities. A detailed analysis of these subtleties and further discussion can be found in Ref.~\cite{Acevedo:2020gro,Bell:2020lmm,Bell:2020jou}.
Note that the estimates presented above neglect the self-interactions of DM particles. Accounting for such self-interactions may in some cases significantly changes (enhances) the capture rate \cite{Guver:2012ba}.

\subsubsection{Capture at star formation}
\label{sec:capture-formation}

In addition to the DM accumulated during the lifetime, stars capture some
amount of DM in the process of their formation. This capture mechanism is
different from the accumulation considered in the previous section as its
first stage is gravitational and thus is independent of the DM properties. We
summarise here this mechanism following
Refs.\cite{Capela:2012jz,Capela:2014ita}. 

Stars are formed from the pre-stellar cores. These baryonic overdensities
create dips in the gravitational potential filled with slow DM particles bound
to the pre-stellar core. These slow particles are, of course, only a small
fraction of the ambient DM density, and their mass density is completely
negligible to that of the baryons. When the star forms, this bound DM is
dragged along by the time-dependent gravitational field of contracting baryons.
This process is slow in the sense that the star formation time, being
determined by the energy dissipation by the baryons, is much larger
than the free fall time. It can therefore be described in the adiabatic
approximation. When the star is finally formed, the space distribution of the
bound DM becomes peaked at the star position. This is the first stage of the DM capture. 

Among the DM particles that form the resulting cuspy profile, of interest to
us are those that have trajectories crossing the newly-born star. These
particles are in the same conditions as the ones captured by direct scattering
(cf. Sect.~\ref{sec:capture-lifetime}): they will sooner or later scatter
off the star nucleons, lose their energy and settle inside the star.

The amount of DM that is captured in this way is proportional to the
density of DM that was gravitatonally bound to the pre-stellar core, and
depends on the parameters of the DM distribution after the adiabatic
contraction of baryons. The bound DM density can be estimated in terms of the
depth of the gravitational potential of the pre-stellar core $\phi$ as
follows: 
\begin{equation}
\rho_{\rm bound} = \rho_{\rm DM} 
{4\pi\over 3} \left( {3\phi\over \pi \bar v^2} \right)^{3/2},
\label{eq:rho_bound}
\end{equation}
where $\rho_{\rm DM}$ is the ambient DM density and its velocity distribution
is assumed to be Gaussian with the characteristic velocity $\bar v$.  The
gravitational potential in turn can be estimated in terms of the parameters of
the pre-stellar core and is in the range $3\times 10^{-12} - 3\times 10^{-11}$
for stellar masses $1-10$ $M_\odot$ \cite{Capela:2012jz}. In all cases one has 
$\rho_{\rm bound}/\rho_{\rm DM} \ll 1$. 

The distribution of the bound DM after the contraction of baryons can  be
calculated numerically. The simulation is straightforward due to two
simplifications: the contribution of the DM into the gravitaional potential is
negligible, and therefore particle trajectories may be simulated one at a time,
and the process is adiabatic so that the detailes of the baryonic contraction
are irrelevant as long as it is sufficiently slow (in practice, several times
slower than the free fall time is enough). Combining the results of 
Refs.~\cite{Capela:2012jz,Capela:2014ita} one obtains the following estimates
of the total mass of the DM that acquires orbits crossing the star and may
eventually be captured, for two different star masses: 
\begin{eqnarray}
&& \text{$M_\odot$:} 
\label{eq:adiabatic_Msun}
\quad \ghost{15pt} M_{\rm tot} \sim 
2\times 10^{-13} M_\odot 
\left({\rho_{\rm DM}\over 100 \, {\rm GeV/cm}^3 }\right) 
\left({10\, {\rm km/s}\over \bar v} \right)^3, 
\\
&& \text{$10~M_\odot$:} 
\label{eq:adiabatic_10Msun}
\quad \ghost{15pt} M_{\rm tot} \sim 
6\times 10^{-11} M_\odot 
\left({\rho_{\rm DM}\over 100\, {\rm GeV/cm}^3 }\right) 
\left({10\, {\rm km/s}\over \bar v} \right)^3. 
\end{eqnarray}
These numbers can be compared to eq.~(\ref{eq:F_Sun_dwarf_lm}). 
Taking $f\sim 10^{-10}$ and assuming for a Sun-like  star lifetime $\sim
10$~Gyr one has from direct capture, eq.~(\ref{eq:F_Sun_dwarf_lm}), $M_{\rm
  tot}\sim 3\times 10^{-15}M_\odot$ for the same environment of a typical
dwarf galaxy. We see that capture at the stage of star formation may give a
dominant contribution to the total amount of accumulated DM.

\subsubsection{Thermalization of captured DM}
\label{sec:therm-capt-dm}

Both mechanisms that have been described above require that DM particles which
have been captured on bound star-crossing orbits interact with the stellar
nucleons in order to eventually settle inside the star. This thermalization process
takes time that is controlled by the DM parameters: the mass $m$ and the
cross section $\sigma$ of DM with nucleons. The thermalization time can be
estimated as follows \cite{Kouvaris:2010jy}. One has to consider separately
two stages. At the first stage, the DM trajectories are mostly outside of the
star while crossing it twice per period. At the second stage the DM particles move
on orbits completely contained inside the star. In both cases the energy loss
happens over many orbits and can be described by a differential equation in a
continuous energy loss approximation. Solving this differential equations gives
the thermalization times for the two stages. For the second stage the initial
conditions are the same for both capture mechanisms considered previously, and one has 
\begin{equation}
t_2 = {m^{3/2}\over \sqrt{3} \rho_* \sigma} {1\over \sqrt{kT_*}} \sim
64 {\rm yr}\, 
\left({m\over 100\, {\rm GeV}} \right)^{3/2} 
\left( {10^{-36}\, {\rm cm}^2\over \sigma  }\right),
\label{eq:therm_t2}
\end{equation} 
where $T_*$ is  the star core temperature which determines the final particle energy. 

For the first stage the initial conditions are different for the two capture
mechanisms. In the case of direct capture they are set by the energy loss in
the first collision. One has \cite{Kouvaris:2010jy}:
\begin{equation}
t_1 \sim  {3\pi m R_*^{3/2} \sigma_{\rm cr} \over 
4m_p c R_g^{1/2} \sigma } \sqrt{\left|{E_*\over E_0} \right| }
\sim 20 {\rm yr}\, 
\left({m\over 100\,{\rm GeV}} \right)^{3/2} 
\left( {10^{-36}\, {\rm cm}^2\over \sigma  }\right),
\label{eq:therm_t1}
\end{equation} 
where $R_*$ is the stellar radius, $R_g$ is its gravitational radius, $m_p$ is
proton mass, $E_*=GMm/R_*$ is the particle binding energy at the star surface,
and $E_0=  m_p c^2 R_g/R_*$ is a typical energy loss in the first collision, so
that $E_*/E_0\sim m/m_p$. 
In the case of the direct capture by a NS the parameters of the NS should be used in these estimates. Assuming $\sigma \sim \sigma_{\rm cr}$, eqs.(\ref{eq:therm_t2}) and
(\ref{eq:therm_t1}) then give $t_1\sim 0.16$~s, $t_2\sim 100$~s for $m=100$~GeV, i.e., in this case the thermalization time is negligibly short. 

For the capture at formation the initial conditions are determined by the size
of the pre-stellar core which sets the size of largest orbits. The ratio of
the final to initial energy becomes instead $E_*/E_0 \sim R_{\rm psc}/R_*$
where $R_{\rm psc}$ is the radius of the pre-stellar core. At $m=100$~GeV this
gives an extra factor 100 in the thermalization time, so that 
\begin{equation}
t_1 \sim  2\times 10^3 {\rm yr}\, 
\left({m\over 100~{\rm GeV}} \right) 
\left( {10^{-36} {\rm cm}^2\over \sigma  }\right).
\label{eq:therm_t1a}
\end{equation} 
The thermalization time becomes much larger than the age of the Universe for
allowed spin-independent cross sections and DM masses in the $100$~GeV range. 

The above estimates of the thermalization time have to be modified in the case
of a direct capture by the NS. In this case the degeneracy of the
nuclear matter has to be taken into account. For heavy DM the modifications
are not dramatic \cite{Goldman:1989nd,Kouvaris:2010jy} because in the case of
a neutron star DM particles falling onto the star have semi-relativistic
velocities, so that for particles much heavier than protons the momentum
transfer is of the order of the neutron Fermi momentum. For lighter DM
particles the modifications are large. The recent detailed calculations can be
found in Refs.~\cite{Bertoni:2013bsa,Cermeno:2017xwb,Garani:2018kkd,Garani:2020wge}. In case of very small interaction cross-sections, $\sigma\ll 10^{-36}~{\rm cm}^2$, and corresponding long thermalization times, the capture process could be affected by influence of external perturbers which  would prevent some particles  from moving into completely contained orbits \cite{Montero-Camacho2019}.

Once the DM particles thermalize with the star, they form a spherical cloud in
the center of a radius 
\begin{equation}
r_{\rm th} = \left({9T_*\over 8\pi G \rho_* m } \right)^{1/2}. 
\label{eq:r_th}
\end{equation} 
For a Sun-like star with a core temperature of $T_*=1.5\times 10^7$~K, core
density $\rho_*=150$~g/cm$^3$, and DM mass of $100$~GeV this cloud is of the
size of $\sim 1000$~km. For a typical old NS the thermal radius is
$r_{\rm th} \sim 20$~cm at temperature $10^5$~K.

\subsection{Signatures of DM in compact stars}
\label{sec:dm-compact-stars}

We have seen in the previous section that, even though the DM may accumulate
in stars, its total amount may only constitute a tiny fraction $\lesssim
10^{-10}$ of the stellar mass. It may nevertheless produce a number of
potentially observable signatures, particularly in compact stars. We discuss
below two of these signatures: the heating of NSs by DM annihilations,
and destruction of the star by DM collapse into a small BH inside the
star. 

\subsubsection{Annihilation and heating}
\label{sec:annihilation-heating}

Because of a very long DM lifetime the decay of the accumulated DM can be
safely ignored. In many conventional DM models, however, the DM particles can
annihilate into the standard model ones. The annihilation rate is proportional to
the square of the DM density which is many orders of magnitude higher for DM
accumulated inside the star compared to the Galactic DM density. The
annihilation may therefore be efficient even for very low annihilation cross
sections, particularly for NSs where the accumulated DM concentrates
in a very small central region. These annihilations produce heat. In the case
of a Main sequence star this heat is smaller compared to energy release of  nuclear reactions by
at least several orders of magnitude (cf. eqs.~(\ref{eq:F_Sun_MW}) --
(\ref{eq:F_Sun_dwarf_lm}) with $f\ll 1$) and is therefore negligible. On the
contrary, NSs have no major internal heat source; the heat produced by DM
annihilations can in principle be detectable
\cite{Kouvaris:2007ay,Kouvaris:2010vv}. 

In principle, depending on the structure of the dark sector, DM annihilations amplified  by its concentration in neutron stars may produce more specific signatures. For instance, it has been pointed out in Ref.~\cite{Leane:2021ihh} that if DM annihilates into sufficiently long-lived particles that can escape from the neutron star and decay outside into the visible sector, the existing experiments such as Fermi and HESS may impose constraints on the DM parameters several orders of magnitude better than those from the direct detection experiments. Here we concentrate on the more model-independent heating signature. 

In case of annihilation, the current amount of DM in a star is determined by
the balance between the accretion and annihilation (for DM masses larger than
$\sim 10$~GeV that we consider here, the evaporation is negligible; for
smaller masses see Ref.\cite{Garani:2021feo}). In this case the number of DM
particles in a star $N(t)$ is governed by the equaion 
\begin{equation}
{dN(t) \over dt} = F - C_A N(t)^2,
\label{eq:annihilation}
\end{equation}
where as before $F$ is the accretion rate and $C_A = \langle \sigma_A v\rangle
/V$ is the thermal averaged annihilation cross section $\sigma_A$ divided by
the annihilation volume $V$. The solution to this equation asymptotes to a
constant over the time scale $\tau =1/\sqrt{FC_A}$. This time scale is
typically very short: for a characteristic value of $\langle \sigma_A v\rangle\sim 10^{-26}$cm$^3$/s and the rate of eq.~(\ref{eq:F_NS}) taken at $f=1$ one finds $\tau \sim
10^{-4}$~yr. For the cross sections as low as $\langle \sigma_A v\rangle \gtrsim 10^{-46}$cm$^3$/s this timescale is still below 1~Myr \cite{Kouvaris:2010vv}. 
In the equilibrium when the annihilation and capture rates
exactly balance each other, the power produced by the annihilation is given
simply by the mass accretion rate of eq.~(\ref{eq:F_NS}).

In the absence of other heat sources, the ultimate NS surface
temperature is determined by equating the mass capture rate (\ref{eq:F_NS}) to
the black body radiation from the star surface $L = 4\pi R_*^2 \sigma_B T^4$
where $\sigma_B$ is the Stephan-Boltzmann constant. This gives
\begin{equation}
T = \left( {mF\over 4\pi R_*^2 \sigma_B} \right)^{1/4}
\sim 5000\,{\rm K}.
\label{eq:T_asymp}
\end{equation}
In the absence of annihilations a NS would cool down below this
temperature in about $2\times 10^7$~yr. Having observed an old NS
with surface temperature in or below this range would thus impose constraints
on the DM annihilation cross sections. Depending on the observed NS temperature and on the DM density at the star location, cross sections a few orders of magnitude below current direct detection limit may be constrained in this way. 

An important observation has been made in Ref.\cite{Baryakhtar:2017dbj} where it was noted that capture of DM particles by the NS {\em itself} provides a heat source comparable in power to that of DM annihilations. The gravitational potential inside the NS is $0.2-0.3$, so capturing a DM particle releases a sizeable fraction of its rest mass in the form of thermal energy. This energy source is therefore only a factor of a few less efficient than annihilations. This mechanism, referred to as 'kinetic heating', is completely free from the assumptions about the DM annihilation cross section. The constraints on the DM from the kinetic heating have been worked out in detail in  Refs.~\cite{Raj:2017wrv,Bell:2018pkk,Acevedo:2019agu,Joglekar:2020liw,Acevedo:2020gro}.  

Internal heating can be important also for WDs. This was recently analysed in \cite{2020IJMPD..2950058P}. These authors studied the case of relatively light weakly interacting massive particles thermalized in compact objects and compared theoretical predictions on additional luminosity with observations of 10 WDs in a globular cluster M4. In the mass range  $\sim 0.1$~--5~GeV the limit on the cross section of DM particles on baryons is $\sigma_{\mathrm{DM-n}}\lesssim$~few~$10^{-41}$~cm$^{2}$. 

A more complicated model of DM annihilation (with formation of metastable mediators decaying into photons) in application to WDs from the same globular cluster M4 was presented in \cite{2018PhRvD..98f3002C}. Mediators can decay to $\gamma$-photons outside a WD, thus this emission can be detected. If mediators decay inside a WD, then an additional contribution to its thermal luminosity appears. Both variants were discussed in  \cite{2018PhRvD..98f3002C} and compared with limits on the thermal luminosity of WDs in M4. This allows to put some model-dependent constrains on the lifetime of mediators.

In principle, the arguments based on the NS heating by DM can be inverted and used to search for DM. In practice, however, the discovery potential of this method is limited by potential presence of alternative heat sources in old NSs \cite{Gonzalez:2010ta,Hamaguchi:2019oev}. Other DM-related effects may help to overcome this difficulty \cite{Casanellas:2010he,Camargo:2019wou,Graham:2015apa,Graham:2018efk,Bell:2019pyc,Acevedo:2019gre,Janish:2019nkk}.

\subsubsection{BH formation and star destruction}
\label{sec:bh-formation-star}

Another potential signature of DM accumulation in stars is related to its
possible collapse into a small seed BH inside the star with the
subsequent accretion of the stellar matter and the star destruction. In order to collapse, the DM
has to be concentrated in a small volume. Only models with non-annihilating DM (e.g., asymmetric DM models \cite{Petraki:2013wwa,Zurek:2013wia}) are relevant in this context as even a tiny annihilation cross section is enough to burn out the accumulated DM. The size of the DM cloud inside the
star depends on the stellar core density and is many orders of magnitude smaller
for compact stars than for Main sequence ones. We concentrate here on NSs and WDs as providing most favorable conditions for the collapse.

The arguments based on the DM collapse into a BHs inside compact stars
can be used in both directions: an observation of a NS or a WD in a
given DM-rich environment would exclude/constrain those DM models/parameters
for wich a BH should have been formed inside a WD/NS and converted it
into a (sub)solar mass BH. Alternatively, in models where this
destruction could take place only in some particularly favorable conditions
one predicts the existence of (sub)solar mass BHs which are not formed
by conventional stellar evolution mechanisms.

The general stages of the DM collapse inside a star are as follows. The
captured DM forms the cloud of radius given by eq. (\ref{eq:r_th}) in the center of the
star. As the DM accumulates, its mass density grows and at some point becomes
larger than the density of the stellar matter. At this point the cloud becomes
self-gravitating and starts to collapse. As the cloud becomes stronger bounded,
the released energy is evacuated through the collisions with the star
nucleons. The shrinking continues either until the extra source of pressure
arises that can stabilize the DM sphere, or until the BH is formed in
the absence of such a source.

At this last stage, the details of the collapse depend on the DM nature --
whether it is bosonic or fermionic. In case of free fermions the Fermi pressure
will halt the collapse unless the number of particles is sufficiently large.
The required number of DM particles is given by an analog of the famous
Chandrasekhar condition which in the case at hand reads:
\begin{equation}
N_{\rm min,f} \sim \left({M_{\rm Pl}\over m} \right)^3 = 2\times 10^{51} 
\left( {100\,{\rm GeV} \over m }\right)^3.
\label{eq:Nmin-fermions}
\end{equation}
Note that for $m\sim 1$~GeV this equation roughly
reproduces the number of neutrons in a NS. In case of
free  bosons this is the quantum uncertainty principle that halts
the collapse if the number of particles is smaller than
\begin{equation}
N_{\rm min,b} \sim \left({M_{\rm Pl}\over m} \right)^2 = 2\times 10^{34} 
\left( {100\,{\rm GeV} \over m }\right)^2.
\label{eq:Nmin-bosons}
\end{equation}
The required number of particles in this case is much smaller than in the case
of fermions. Note, however, that eq.~(\ref{eq:Nmin-bosons}) assumes that bosons are non-interacting; interactions may significantly modify this equation as will be discused below.

Consider first the case of free fermions. In this case the number of DM particles required to make a BH is given by eq.~(\ref{eq:Nmin-fermions}). Let us compare it to the number that can be accumulated according to
eq.~(\ref{eq:F_NS}). Assuming the NS lifetime of 1~Gyr,
eq.~(\ref{eq:F_NS}) implies the total accumulated amount of $8\times
10^{42}$~GeV in the conditions typical for Milky Way, regardless of the DM
mass. Dividing by the mass and requiring that the resulting number is larger
than that of eq.~(\ref{eq:Nmin-fermions}) one finds the condition 
\begin{equation}
m \gtrsim 10^7~{\rm GeV} ,
\label{eq:exclusion_fermions}
\end{equation}
in agreement with the results of Ref.~\cite{Goldman:1989nd}. 
The DM models with heavier fermionic DM would then contradict to observations
of NS in the Milky Way, provided the DM has scattering cross section on
neutrons that is larger than $\sigma \gtrsim 10^{-45}$~cm$^2$ so that
eq.~(\ref{eq:F_NS}) is valid with $f=1$. 

This constraint on the mass can be improved by excluding lower DM masses in
two ways. The minimum mass to form a BH scales like the square root of the
capture rate. The latter, in turn, scales like $\rho_{\rm DM}/\bar
v$. Observation of NSs in DM-rich environments thus leads to
stronger constraints. For instance, in dwarf galaxies both the DM density is
higher by a factor up to $\sim 200$, and the velocity of DM particles is lower
by up to $30$ times. Thus, in dwarf galaxies the capture rate may be higher by
nearly 4 orders, and correspondingly the constraints on the fermionic DM mass
start at masses lower by 1.5-2 orders of magnitude compared to
eq.~(\ref{eq:exclusion_fermions}). The second improvement may be achieved by
taking in consideration the accumulation of DM by the NS progenitor. The
efficient accumulation of DM in the Main sequence stars requires large
DM-nucleon cross section in view of eqs.~(\ref{eq:sigma_cr}). The argument,
therefore, applies to models with the spin-dependent DM-nucleon cross section
that is less constrained. Depending on the value of this cross section, masses
as low as $\sim 10$~TeV can be potentially constrained \cite{Kouvaris:2010jy}.

We have considered so far non-interacting fermions. Adding self-interactions may change the collapse conditions. The simplest Yukawa interaction with the scalar mediator gives the attractive interaction which may reduce the number of DM particles required for the collapse  \cite{Kouvaris:2011gb,Bramante:2013nma,Kouvaris:2018wnh}. Note, however, that the non-relativistic approximation becomes inadequate in the relativistic regime close to the collapse \cite{Gresham:2018rqo}, which may limit this reduction to a mere factor of a few as compared to the non-interacting case. 

Let us turn now the case of bosonic DM. In the case of bosons one more phenomenon has to be taken into account, the formation of the Bose-Einstein condensate
(BEC) \cite{Kouvaris:2011fi}. The condition for the BEC formation (large phase
space density) can be formulated in terms of DM number density $n$ and the
star core temperature $T_c$ and reads
\begin{equation}
n \gtrsim 5\times 10^{28} {\rm cm}^{-3} 
\left({m\over {\rm GeV}}\right)^{3/2}
\left({T_c\over 10^5{\rm K}}\right)^{3/2},
\label{eq:BEC_cond}
\end{equation}
where we assumed that DM particles are in thermal equilibrium with the star. 
In view of eq.~(\ref{eq:F_NS}) this condition becomes satisfied very
quickly. Once the BEC is formed, newly captured DM particles go into the
condensed state. The size of the condensate $d$ is determined by the size of the lowest quantum level of DM in the gravitational potential of the NS: 
\begin{equation}
d = \left({8\pi \over 3} G\rho_c m^2 \hbar^{-2} \right)^{-1/4}
\sim 2\times 10^{-4} {\rm cm} \left({{\rm GeV}\over m}\right)^{1/2},
\label{eq:BEC-size}
\end{equation}
where $\rho_c$ is the star core density. This size is much smaller than the
size of the thermal DM cloud. 
Because of the small size of the BEC, the density of DM in the BEC quickly
exceeds the density of nucleons in the star. This happens when the accumulated
DM mass becomes larger than 
\begin{equation}
M\gtrsim 10^{28}{\rm GeV} \left({m\over {\rm GeV}}\right)^{-3/2}, 
\label{eq:BEC-selfgrav}
\end{equation}
i.e., almost immediately after the star formation.  The self-gravitating BEC then grows until the condition of BH formation is satisfied.

In the case of zero DM self-interactions there is nothing that can stop the BH
formation once the number of DM particles exceeds that given in
eq.~(\ref{eq:Nmin-bosons}). In this case the constraints can be imposed on the
DM parameters. Requiring that the BH that is formed is heavy enough to grow by
accretion faster than it evaporates by the Hawking radiation, one may exclude
the DM masses in the range $100{\rm keV} - 10$~GeV for DM-nucleon cross
sections at the current experimental limit \cite{Kouvaris:2011fi,McDermott:2011jp}. For heavier masses the BH formed in the DM collapse is too light and evaporates through Hawking radiation faster than it accretes the matter of the star \cite{Kouvaris:2012dz} with no observable consequences apart from heating of the star as already discussed above. 
Adding even a tiny $\lambda \phi^4$ self-interactions shifts the exclusion region into the multi-TeV range (see fig.2 of \cite{Kouvaris:2011fi}). 
A detailed recent analysis of these constraints can be found in \cite{Garani:2018kkd}. 

It has been pointed out in Ref.\cite{Bell:2013xk} that self-interactions {\em must} be present at a non-negligible level, the reason being as follows. Even if a direct coupling between DM particles is not included in the Lagrangian, the DM interactions with nucleons that are necessary for capture will induce the DM-DM self-interactions in the higher orders of perturbation theory. There is however, a subtlety in this argument. The DM-nucleon scattering which is necessary for capture occurs at low energies, while the DM-DM
self-interactions present an obstacle for collapse at very high (Planckian)
values of fields. These two scales are not necessarily related. So, while it
is easy to avoid the constraints by adding an appropriate DM self-interaction,
the assumption that self-interactions are negligible at the collapse is
{\em not} in contradiction with the non-zero DM interaction with nucleons at
low energies. 

Having excluded some regions of the DM parameter space by NS implosions due to formation of seed BHs automatically implies, by continuity, that for the parameters close to those regions but not excluded, one may look for signatures of these implosion events. These signatures may include particular gravitational wave events \cite{Kurita:2015vga,Bramante:2017ulk,Takhistov:2017bpt,Nelson:2018xtr}, quiet kilonovae \cite{Bramante:2017ulk}, supernovae \cite{Bramante:2014zca,Bramante:2015dfa,Bramante:2015cua}, an impact on star population \cite{Casanellas:2011qh}, and existence of (sub)Solar-mass BHs \cite{Kouvaris:2018wnh,LIGOScientific:2019kan} not expected to result from conventional stellar evolution mechanisms.

\section{Axions and neutron star magnetospheres}

\subsection{Theory}
One of the unsolved issues  of the Standard model (SM) of particles is strong CP problem, i.e. absence of CP-violation in the quantum chromodynamics (QCD). The theory contains terms that could generate CP-violation, so extreme closeness of parameter $\theta$ that describes strength of the violation and could take values  anywhere in $[0,2\pi]$ interval, to zero seems unnatural. 
One of the most elegant solutions of the strong CP problem was suggested by Peccei and Quinn (\cite{PQ})  --- to treat $\theta$ as a new dynamical field, rather than a fixed parameter. Dynamics of this field   could drive the value of CP-parameter to zero, effectively solving the problem. There also would be particles corresponding to this field --- axions \cite{Wilczek1978, Weinberg1978}.
There is no unique way to  add axion to the SM, and two most popular models are KSVZ (Kim-Shifman-Vainshtein-Zakharov) and DFSZ (Dine-Fischler-Srednicki- Zhitnitsky) \cite{Kim1979,Shifman1980,Zhitnitsky1980,Dine1981} which predict  couplings of axion to SM particles of similar strengths and frequently serve as benchmarks.  In these models the axion mass $m_a$ is a free parameter that can change in a very wide range and should be found from experiments. One of the general prediction for axions is that their coupling constants, e.g. the coupling constant to photons 
$\gag $, are proportional to  mass $m_a$, meaning that less massive axions become progressively more elusive for direct or indirect detection. Still, it is at least theoretically possible to reach this line, or, given uncertainties, band on $m_a-\gag$ plane for some values of $m_a$ and therefore test model of QCD axion in a  certain  mass range.
In low-energy effective field theories derived from some string models more general class of axion-like particles (ALPs) could emerge (e.g., \cite{Svrcek2006,Choi2009,Arvanitaki2010}). For ALPs  mass and coupling constants are independent.

Axions (and ALPs) are also regarded as a possible DM candidate \cite{Preskill1983, Abbott1983}. Though they are very light by standards of WIMPs, with masses $<1~$eV, still axions could contribute to cold DM (CDM). A  concise review of axion and ALPs and present  status of different experiments searching for them could be found in \cite{pdg_2020}. Properties of axions and ALPs which are relevant to us  are very close, so from now on we will refer to both of them as axions.  Throughout this section we use units with $\hbar=c=1$ and $\alpha=e^2/4\pi\approx1/137$.

Most of the search methods rely on the very weak axion coupling to photons. The lagrangian of axion-photon system in vacuum without QED corrections is described as follows:
\begin{equation}
\mathcal{L}=-\frac{1}{4} F_{\mu\nu} {F}^{\mu\nu}+\frac{1}{2}\left(\partial_{\mu}a\partial^{\mu}a-m_a^2a^2\right)+\frac{1}{4}\gag  a F_{\mu\nu} \widetilde{F}^{\mu\nu},
 \label{eq:ax_lagrangian}
\end{equation}
where $F_{\mu\nu}\equiv \partial_{\nu}A_{\nu}-\partial_{\mu}A_{\mu}$ is the electromagnetic field tensor, $\widetilde{F}^{\mu\nu}\equiv \frac{1}{2}\epsilon_{\mu\nu\rho\sigma}F^{\rho\sigma}$ is its dual, and $a, m_a$ are the axion  field and mass respectively.
The coupling term could be rewritten using magnetic and electric fields strengths $\vec{E}, \vec{B}$:  $\frac{1}{4}\gag  a F_{\mu\nu} \widetilde{F}^{\mu\nu}= \frac{1}{4}\gag  a \vec{E}\cdot\vec{B}$.
The equations of axion electrodynamics could be derived from the Lagrangian (\ref{eq:ax_lagrangian})  \cite{Wilczek1987,Battye2020}:

\begin{align}
\nabla\cdot\vec{E}=-\gag\vec{B}\cdot\nabla a ,
\label{eq:ax_maxwell_divE}\\
\nabla\times\vec{E}=-\frac{\partial \vec{B}}{\partial t},
\label{eq:ax_maxwell_rotE}\\
\nabla\cdot\vec{B}=0 ,
\label{eq:ax_maxwell_divB}\\
\nabla\times\vec{B}=\frac{\partial \vec{E}}{\partial t}+\gag \dot{a}\vec{B}+\gag \nabla a \times \vec{E},\label{eq:ax_maxwell_rotB}\\
\label{eq:ax_KGF}
\Box a+ m_a^2a=\gag \vec{E}\cdot\vec{B}.
\end{align}

Let us assume that there is some background magnetic field $\vec{B}_0$ and perturbations $\vec{E},\vec{B}$ propagating over it. In most astrophysical scenarios background electric field $\vec{E}_0$ could be safely set equal to zero. Furthermore, as we deal with propagating EM waves, it is sufficient to describe  $E$-component only: 

\begin{align}
\Box \vec{E} +\nabla(\nabla\cdot\vec{E})= -\gag \ddot{a}\vec{B}_0.\label{eq:EM_wave}\\
\Box a+ m_a^2a=\gag \vec{E}\cdot\vec{B}_0.\label{eq:ax_KGF_2}
\end{align}

It is illuminating to consider the simplest setup: the EM wave and axion propagate along $z$-axis in a static and uniform background magnetic field $\vec{B}_0$. As the coupling is sourced by the scalar product $\vec{E}\cdot\vec{B}_0$, only transverse  component  of the magnetic field $\vec{B}_{0}$  enters  into the equations of motion --- we use it to define the direction of $x$-axis. Also,   only one polarization $E_{||}=E_x$ couples to axion and experiences conversion to axion and vice versa, as for the other one the scalar product $\vec{E}\cdot\vec{B}$ vanishes.

Factorizing out temporal dependence $\sim e^{i\omega t}$, we arrive at the following set of equations:

\begin{align}
\begin{pmatrix}
    \omega^2+\partial_z^2-m_a^2        & \omega\gag B(z)  \\
    \omega\gag B(z)       & \omega^2+\partial_z^2
\end{pmatrix}
\begin{pmatrix}
    a  \\
    E_{||}/\omega
\end{pmatrix}=0
\label{eq:eq_motion}
\end{align}

These second-order  equations could be solved numerically. Another approach is to linearize them and obtain analytic results from first-order equations.  The details of the linearization procedure depends on the dispersion relation for axion $\omega^2=m_a^2+k^2$, where $k$ is the wave vector, and slightly differs for cases of relativistic and non-relativistic axion. The former  case, which is e.g. relevant for propagation of X-ray photons  in magnetospheres  of NS, was  first thoroughly studied in \cite{Raffelt1988}. For ultrarelativistic axions   $k\approx \omega$ and the linearized equations have a very simple form\footnote{For non-relativistic axions an analogous expression can be found e.g. in \cite{Battye2020}.}:
\begin{align}
\begin{pmatrix}
   \omega-i\partial_z-\frac{m_a^2}{2\omega}        &\frac{\gag B(z)}{2}  \\
    \frac{\gag B(z)}{2}        & \omega-i\partial_z
\end{pmatrix}
\begin{pmatrix}
    a  \\
    E_{||}/\omega
\end{pmatrix}=0
\label{eq:eq_motion_lin}
\end{align}

These Schrodinger-like equations describe behaviour of a system of two mixed states. We are mainly interested in a probability of conversion between them. In case of homogeneous constant magnetic field $B$ the probability of conversion of one particle into another depends on the travelled distance $L$ \cite{Raffelt1988,Fairbairn2011}:
\begin{align}
P(L)=\frac{\gag^2B^2}{m_a^4/4\omega^2+\gag^2B^2}\sin\left(\frac{L\Delta_{osc}}{2}\right),
\label{eq:p_conversion}\\
\Delta_{osc}^2=m_a^4/4\omega^2+\gag^2B^2.
\label{eq:osc_length}
\end{align}

Photons and axions are in a  strong mixing regime, when $\gag B \gg m_a^2/2\omega$. In this regime  almost full conversion of photons to axions and back is possible with a characteristic length scale of oscillations $l_{osc}=2\pi/\Delta_{osc}=2\pi/\gag B$. In the opposite case, $\gag B \ll m_a^2/2\omega$, the amplitude of oscillations is greatly suppressed. This suppression comes from the fact that the phase velocities of waves corresponding to massless photon and massive axion are slightly different and conversion process has only limited time to build up while these waves propagate in phase. After that the conversion process falls out of the resonance, effectively limiting the amplitude of oscillations.

This picture of axion/EM wave propagating in vacuum could be generalized to a more physical scenario. Now we take into account effects of medium and QED-induced terms which changes  velocity of propagation of EM waves \cite{Raffelt1988, Fairbairn2011}. Equation (\ref{eq:eq_motion_lin}) should be rewritten as follows:

\begin{align}
\begin{pmatrix}
   \omega-i\partial_z+\Delta_a       &\Delta_M  \\
    \Delta_M        & \omega-i\partial_z +\Delta_{||}
\end{pmatrix}
\begin{pmatrix}
    a  \\
    E_{||}/\omega
\end{pmatrix}=0,
\label{eq:eq_motion_lin1}
\end{align}
where $\Delta_a=-\frac{m_a^2}{2\omega}$, $\Delta_M=\frac{\gag B\sin\Theta}{2}$, and $\Theta$ is the angle between the direction of axion propagation and background magnetic field. In astrophysical sources the mixing term $\Delta_{||}$ is a sum of two parts $\Delta_{||}=\Delta_{pl}+\Delta_{QED,||}$. The first one arises due to the influence of plasma, where  photon acquires  effective mass and propagates slower than in vacuum: $\Delta_{pl}=-\frac{\omega_{pl}^2}{2\omega},~\omega_{pl}=\sqrt{\frac{4\pi\alpha n_e}{m_e}}$, $\alpha$ is the fine-structure constant, $n_e$ is the number concentration of electrons  and $m_e$ is the  electron charge \footnote{If the main charge carriers are not electrons, than the $n_e, m_e$ should be substituted for $n_c, m_c$.}
The second term $\Delta_{QED,||}$ comes from QED-induced corrections  and depends on the strength of magnetic field $B$,  $\Delta_{QED,||}=\frac{1}{2}q(b)\omega \sin^2\Theta$, $q(b)$ is a function of parameter $b\equiv\frac{B}{B_{cr}}$,  $B_{cr}=m_e^2/e=4.4\times10^{13}~$G, which can be approximated in the following form \cite{Potekhin2004}:

\begin{equation}
q=\frac{7\alpha}{45}b^2 \hat{q},\, \, \,  \hat{q}=\frac{1+1.2b}{1+1.33b+0.56b^2}.
 \label{eq:q_approx}
\end{equation}

Usually it is sufficient to use the small-$b$ approximation  $\Delta_{QED,||}=\frac{7\alpha}{90}\omega \left(\frac{B}{B_{cr}}\right)^2 \sin^2\Theta$. The only possible exception could arise when one considers  propagation of EM waves in the immediate vicinity of magnetars, where magnetic field could be considerably higher than $B_{cr}$.
Introduction  of these new mixing terms slightly modifies the expression (\ref{eq:p_conversion}) for probability $P(L)$  :  

\begin{align}
P(L)=\frac{4\Delta_M^2}{(\Delta_{pl}+\Delta_{QED,||}-\Delta_a)^2+4\Delta_M^2}\sin\left(\frac{L\Delta_{osc}}{2}\right),
\label{eq:p_conversion_full}\\
\Delta_{osc}^2=(\Delta_{pl}+\Delta_{QED,||}-\Delta_a)^2+4\Delta_M^2.
\label{eq:osc_length_full}
\end{align}

It can be seen that now it is  possible for photons and axions to experience very efficient conversion when the resonance condition
$(\Delta_{pl}+\Delta_{QED,||}-\Delta_a)=0$ is met. This effect is used in radio searches for axions both in laboratory and in observations of compact stars (see  Section \ref{sec:ns_haloscope} below).
On the other hand, as $\Delta_{QED,||}$ and $\Delta_a$ have different signs, QED-induced term can further suppress conversion, comparing to the vacuum case. This is relevant for propagation of X-ray photons and axions of correspondind energies in magnetospheres of MWDs and NSs.

To conclude this section, we will present some benchmark values for the mixing terms, using typical values for relevant  parameters:

\begin{align}\nonumber
\Delta_a=-\frac{m_a^2}{2\omega}=-5\times10^{-16}\left(\frac{m_a}{1~\mu\mathrm{eV}}\right)^2\left(\frac{\omega}{1~\mathrm{keV}}\right)^{-1}~\mathrm{eV}=\\
=-2.538\times10^{-11}\left(\frac{m_a}{1~\mu\mathrm{eV}}\right)^2\left(\frac{\omega}{1~\mathrm{keV}}\right)^{-1}~\mathrm{cm}^{-1},
\\
\Delta_{pl}=-\frac{\omega_{pl}^2}{2\omega}=-\frac{2\pi\alpha n_e}{m_e\omega}=\nonumber\\=-6.87\times10^{-15}\left(\frac{n_e}{10^{10}~{\mathrm{cm}^{-3}}}\right)^2\left(\frac{\omega}{1~\mathrm{keV}}\right)^{-1}~\mathrm{eV}=\nonumber\\
=-3.48\times10^{-10}\left(\frac{n_e}{10^{10}~{\mathrm{cm}^{-3}}}\right)^{-1}\left(\frac{\omega}{1~\mathrm{keV}}\right)^{-1}~\mathrm{cm}^{-1},
\\
\Delta_{M}=\frac{\gag B\sin\Theta}{2}=9.76\times10^{-10}\left(\frac{\gag}{10^{-10}~\mathrm{GeV}^{-1}}\right)\left(\frac{B}{10^{12}~\mathrm{G}}\right)\sin\Theta~\mathrm{eV}=\nonumber\\
=4.94\times10^{-5}\left(\frac{\gag}{10^{-10}~\mathrm{GeV}^{-1}}\right)\left(\frac{B}{10^{12}~\mathrm{G}}\right)\sin\Theta~\mathrm{cm}^{-1},
\\
\Delta_{QED,||}=\frac{7\alpha}{90}\omega \left(\frac{B}{B_{cr}}\right)^2 \sin^2\Theta=\nonumber\\=9.33\times10^{-5}\left(\frac{B}{10^{12}~\mathrm{G}}\right)^{2}\left(\frac{\omega}{1~\mathrm{keV}}\right)\sin^2\Theta~\mathrm{eV}=\nonumber\\
=4.73\left(\frac{B}{10^{12}~\mathrm{G}}\right)^{2}\left(\frac{\omega}{1~\mathrm{keV}}\right)\sin^2\Theta~\mathrm{cm}^{-1}.
\label{eq:mixing_terms}
\end{align}
Relative strength of these terms defines which effect dominates during propagation. In realistic situations where the field is inhomogeneous the probability is calculated by numerical integration of equations (\ref{eq:eq_motion_lin1})  from some starting point to infinity\cite{Raffelt1988}:

\begin{equation}
P_{a\rightarrow\gamma}=\left|\int\limits_{R_{0}}^{\infty}dr'\Delta_M(r')e^{i\Delta_ar'-i\int\limits_{R_{0}}^{r''}dr''\Delta_{QED,||}(r'')}\right|^2.
 \label{eq:p_conversion_integral}
\end{equation}

\subsection{Hot axions}
Extremely strong magnetic fields around  NSs naturally make them  obvious targets for searches for  signatures  of axion-photon oscillations.  First, it was suggested to search for high-energy X-ray photons from conversion of axions produced in  cores of  NSs \cite{Morris1986,Raffelt1988}. Temperature in central parts of these objects can reach extremely high values, $T_c\sim10^9$~K,  and the  axions could be generated  there in bremsstrahlung process during interactions of nucleons, $n+n\longrightarrow n+n+a$.  These axions would propagate freely in the interiors of the NS, but could  convert in the external magnetic fields into energetic photons that could be easily distinguished from thermal photons coming  from the surface as its temperature is much lower, $T_s\sim10^{6}$~K.
Inverse process of photon to axion conversion could also lead to emergence of potentially observable spectral and polarization features in X-ray and other energy ranges \cite{Lai2006, Chelouche2009, Perna2012}.

At $\mathcal{O}(\mathrm{keV})$ energies conversion takes place in a weak mixing mode almost everywhere, as could be seen from eqs. (\ref{eq:mixing_terms}), and the dominating term is $\Delta_{QED,||}$. Particles  could  still experience photon-axion resonance, when resonance condition $\Delta_{pl}+\Delta_{QED,||}-\Delta_a=0$ is met. This is possible when $\Delta_{pl}\approx-\Delta_{QED,||}$, which defines  corresponding density \cite{Lai2006}:
 
\begin{equation}
 \rho_{res}=2.25\times10^{-4}\frac{1}{Y_e}\left(\frac{B}{10^{12}~\mathrm{G}}\right)^{2}\left(\frac{\omega}{1~\mathrm{keV}}\right)^{2}~\mathrm{g~cm}^{-3},
 \label{eq:resonance_density}
\end{equation}
where $Y_e$ is the electron fraction. In resonance photon-axion system is in the strong mixing regime and it is theoretically possible to obtain complete conversion from one state to another. However, the needed oscillation length $l_{osc}=\frac{\pi}{\Delta_M}$ is much larger than the characteristic scale of the NS atmosphere $h_{atm}\sim\mathcal{O}(10)$~cm, let alone more narrow region where density is close to the value which makes the  resonance possible. Because of that the contribution from the resonance conversion is subdominant and in all calculations of conversion probability it is possible to neglect the plasma term $\Delta_{pl}$.

Hot axions generated in the centers of NSs  have a modified thermal distribution \cite{Iwamoto1984,Nakagawa1988} with spectrum peaking around $E_{peak}\sim3.3T_c$ and for temperatures $T_c\sim10^{9}$~K the flux is  non-negligible in 10-1000 keV energy range.
It is customary to calculate probability of conversion using eq. (\ref{eq:p_conversion_integral}) assuming a simple dipolar model for a NS magnetic field and purely radial propagation \cite{Fortin2018, Buschmann2021}. This probability could be estimated as follows \cite{Buschmann2021}:

\begin{align}
P_{a\rightarrow\gamma}=6\times10^{-3}\left(\frac{\gag}{10^{-10}~\mathrm{GeV}^{-1}}\right)^2\left(\frac{B}{10^{12}~\mathrm{G}}\right)^{2/5}\nonumber\\\left(\frac{\omega}{1~\mathrm{keV}}\right)^{-4/5}\left(\frac{R_{NS}}{10~\mathrm{km}}\right)^{6/5}.
 \label{eq:p_conversion_estimate}
\end{align}

Multiplying  theoretically expected axionic spectrum  by the probability it is now possible to obtain spectrum of X-rays from the conversion and compare it to observations. Also polarization measurements could be used to constrain properties of axions: converted photons could have only single  polarization $E_{||}$, also known as O-mode (ordinary), while many models of NS atmospheres predict that NS emission, especially at lower energies $<1$~keV, is primarily polarized in perpendicular X-mode (extraordinary) \cite{vanAdelsberg2006} and an admixture of differently polarized mode could be in principle detected \cite{Fortin2019}.

Different classes of NSs was suggested to use for this type of searches. First, the magnetars \cite{Fortin2018,Fortin2019} --- NSs with extremely high magnetic fields exceeding $10^{14}$~G and high  central temperatures, up to $\sim10^{9}$~K  which boosts their axionic luminosities. There are obvious downsides as well: all of them are rather distant and that greatly decreases the flux observed at the Earth. Also, these objects demonstrate high-energy activity which makes it non-trivial to disentangle possible contribution from axion-photon oscillations. Nevertheless, the observations of magnetars in the energy range 10-1000 keV  allowed to put meaningful constraints on the product of couplings $\gag\times g_{ann}$ \cite{Lloyd2021,Fortin2021}. The most stringent constraints come from magnetar 1E 1547.0-5408: $\gag\times g_{ann} <4.4\times10^{-20}~\mathrm{GeV}^{-1}$ for $T_c=10^{9}$~K.

Another interesting set of candidates are X-ray dim isolated NSs (XDINSs, also known as the Magnificent Seven) (e.g. \cite{Posselt2007}). These NSs are detected mainly in X-rays (plus, dim optical counterparts are known for most of them) and  their emission is thought to originate at their surfaces with spectra which are  very close to blackbody with temperatures $T_s\sim 0.1~$~keV. Detected evolution of spin periods allowed to infer magnetic fields of these objects: $B\sim10^{13}$~G which makes possible  quite effective conversion. Their relative proximity to the Earth, with distances in 100-600 pc range also benefits searches. Deep observations in 2-8 keV range by \textit{ XMM-Newton} and \textit{Chandra} revealed existence of X-ray excess  for two XDINS: RX J1856.6-3754 and RX J0420.0-5022 \cite{Dessert2020}. This excess could be explained by axion-photon conversion if $\gag\times g_{ann}\sim4\times10^{-20}~\mathrm{GeV}^{-1}$  (only central value is quoted, systematic and statistical uncertainties could exceed an order of magnitude ) \cite{Buschmann2021}. This value is only  slightly lower than one coming from a combination of constraints from CAST helioscope on $\gag$ and from SN 1987A on $g_{ann}$, so it is possible that this result would soon be tested with new experiments such as IAXO.

Photon to axion conversion and induced changes in observed spectrum and polarization of NS X-ray emission was extensively studied in \cite{Perna2012}. Although the possible constraints depend now only on $\gag$ coupling, the theoretical uncertainties in modelling of  NS atmospheres greatly complicate the task of setting them.
Usually it is assumed  that below 1-2 keV intrinsic emission of the NS is X-polarized due to an  increased opacity for O-mode. As X- and O- mode could convert into each other in process similar to axion-photon conversion \cite{Lai2006, Perna2012} the emission is mostly polarized in O-mode at higher energies --- originally produced X-polarized photons experience conversion into O-photons in so-called vacuum resonance. Vacuum and axion-photon resonances occur at well separated regions, so it is possible to treat them independently and instead of three-state system (X, O, axion) study consequently two two-state system (O, axion) and (X, O) \cite{Lai2006} 
It was shown that if the photons come from hot spots rather than entire surface, their conversion to axions would lead to phase-dependent suppression of high-energy tail of the spectrum \cite{Perna2012}. In the opposite case, the conversion would affect the inferred radius of NS which could lead to contradictions with values found by other methods.

Another possible way to use photon to axion conversion was suggested recently in \cite{Zhuravlev2021} --- optical photons emitted from the surface of XDINS should also follow blackbody distribution with temperature that can be inferred from X-ray observations of these objects. Conversion into axions would lead to dimming in the optical part of the spectrum which theoretically allows to probe coupling constraints down to values $\gag\sim10^{-11}~\mathrm{GeV}^{-1}$.

Further observational progress would be reached with new generation of X-ray telescopes: \textit{Athena} \cite{ATHENA} would reach the level of sensitivity which would make possible phase-resolved observations of  XDINs and magnetars. Effectiveness of the axion-photon conversion process varies with rotational phase due to changing orientation of magnetic field, so these observations could make crucial contribution to the field. Also polarimetric  observations with telescopes like \textit{XIPE} \cite{XIPE} would significantly further our progress in understanding of NS atmospheres and  propagation of X-ray photons in them, thus allowing to put stronger constraints  on axion properties.

\subsection{Compact object as haloscopes}
\label{sec:ns_haloscope}

Another promising method of axion detection was initially suggested in \cite{Pshirkov2009}. If the axions are the main  DM component, than their conversion in strong magnetic fields of NS magnetosphere could be possibly revealed  by the presence of sharp spectral feature, similar to the laboratory haloscopes used to search for axions (e.g. ADMX \cite{ADMX}). 

Flux of axions through magnetosphere is enhanced by gravitational focusing and infalling particles travelling through the inner regions of the magnetosphere have a chance to convert into photons.  In this case we are dealing with conversion of non-relativistic axions, so the resulting signal would be almost monochromatic at the frequency corresponding to axion mass $m_a$ with certain shift and  widening due to the  Doppler effect.

Renewed interest to this method of axion searches lead to fast progress in our understanding of the  details of the process \cite{Huang2018,Hook2018,Safdi2019,Battye2020,Leroy2020,Witte2021, Battye2021a}.
Conversion process is slightly different for non-relativistic axions: governing linearized equations are altered \cite{Battye2020},  and  conversion occurs in a resonant mode,  in a stark contrast to conversion of 'hot' axions, as discussed above.  Also plasma effects that influence propagation of EM waves are much more important for conversion of non-relativistic axions \cite{Witte2021, Battye2021a}.

 One of the main properties that directly affects potential sensitivity of radio observations is the resulting width of the radio line. It is mostly  defined by interaction of infalling axions with rotating plasma of a NS magnetosphere \cite{Battye2020} and can reach values as high
as:
\begin{equation}
 \frac{\delta f}{f}\sim\mathcal{O}\left(\frac{\Omega r_c}{c}\right)=6\times10^{-4}\left(\frac{\Omega}{1~\mathrm{Hz}}\right)\left(\frac{r_c}{200~\mathrm{km}}\right),
 \label{eq:Doppler}
\end{equation}
where $r_c$ is the critical radius where conversion takes place, $\omega_{pl}(r_c)=m_a$. This contribution due to Doppler broadening strongly dominates over one caused by DM particle velocity dispersion $\bar{v}$: $$\frac{\delta f_{disp}}{f}=\frac{{\bar{v}}^2}{2c^2}= 8 \times 10^{-7} \left(\frac{\bar{v}}{100~\mathrm{km~s^{-1}}}\right)^2.$$
More rigorous treatment of 3D propagation in plasma, including ray-tracing technique, lead to modification of the magnitude of the expected effect, when compared to simple 1D calculations. In 3D non-radial trajectories of photons unlike axions  are not rectilinear and that in turn leads to de-phasing and suppression of resonance which results in somewhat lower flux from conversion \cite{Witte2021}. Also plasma effects   could   lead to emergence of rich temporal structure of a signal.

Due  to their proximity and relatively high magnetic fields XDINSs are again considered to be very good observational targets for searches in radio domain.  Observations of  RX J0720.4-3125 and RX J0806.4-4123 with  Green Bank radio telescope found no excess signal  and that  allowed to set constraints: $\gag<10^{-11}~\mathrm{GeV}^{-1}$ for $5.5<m_a<7.0~\mu eV$ \cite{Foster2020}. 

Signals from NSs could be greatly increased if they reside in regions with enhanced DM  density. Another candidate that potentially could lead to even more stringent constraints is magnetar PSR J1745-2900  with $B\approx1.6\times10^{14}$~G which is only 0.1 pc away from the central supermassive black hole Sgr A$^*$. The boost factor which can be defined as ratio of DM density at the NS position relative to DM density near the Solar system $\eta=\rho_{DM}/\rho_{DM,\odot}$ could reach $1.5\times10^{5}$ in Navarro-Frenk-White(NFW) model of the DM halo and could exceed $10^9$ in models with a spike near the SMBH. This object was extensively studied using Effelsberg observations \cite{Foster2020} and archival data from Karl G. Jansky Very Large Array  \cite{Darling2020a,Darling2020b, Battye2021b}. Constraints can be set in several broad windows from 4 to 160 $\mu$eV, which correspond to radiotelescope frequency bands   spanning from 1 to 40 GHZ. Assuming  NFW model \cite{1997ApJ...490..493N} constraints  can reach $\gag<10^{-11 }~\mathrm{GeV}^{-1}$ level, while they could be two orders of magnitude stronger if there was a DM spike around the SMBH.
At the moment the greatest uncertainty comes from our limited knowledge of DM distribution near the Galactic center. 
In the future, supreme sensitivity of the SKA would allow to test couplings as small as $\gag \sim 10^{-13 }~\mathrm{GeV}^{-1}$ even assuming NFW model \cite{Witte2021}.

MWDs could also be a possible targets of observations, as the conversion of DM could  analogously take place in their magnetospheres. In \cite{Wang2021} it was shown that future observations of MWD 2010+310 with SKA  phase 1 can probe coupling as low as  $\gag\sim10^{-12 }~\mathrm{GeV}^{-1}$ in  0.2-3.7 $\mu$eV mass range.

\subsection{Axions and FRBs}
Fast radio bursts (FRBs) are  extragalactic radio flares with millisecond duration discovered by \cite{2007Sci...318..777L}. The question of their origin  is one of the major astrophysical mysteries nowadays (see reviews e.g. in \cite{Popov2018, 2020Natur.587...45Z, 2021SCPMA..6449501X}). Axion to photon conversion in strong magnetic field was one of the first scenarios involving astroparticle physics. 

Sizable fraction of axions constituting DM could undergo process of condensation, forming relatively dense clusters with masses $\sim\mathcal{O}(10^{-12}~M_{\odot})$ and characteristic sizes of $\sim\mathcal{O}(100~\mathrm{km})$. If  during a passage through magnetosphere of a NS such cluster experiences partial or full conversion and the axion  mass falls in the $\mathcal{O}(\mu\mathrm{eV})$  range,  there would be a bright short radio flare.  Conversion of $\sim\mathcal{O}(10^{-13} -10^{-12}~M_{\odot})$ into EM radiation  could  have explained outstanding energetics of FRBs with  total emitted energy reaching $E_{tot}=10^{40}$~ergs \cite{Tkachev2015, Iwazaki2015}.

As it was quickly demonstrated, this simplified scenario could not work as axion miniclusters are too loosely bound and would be destroyed by tidal forces long before they could reach the region of efficient conversion and so duration of the resulting flare would be of order of seconds rather than milliseconds \cite{Pshirkov2017}.
However, accurate treatment of equations, describing axion condensation shows that there is a branch of solutions, which correspond to much more concentrated   objects, dubbed {\it dense axion stars}. With masses around $10^{-13}~M_{\odot}$ but characteristic sizes less than about a meter they are immune to forces of tidal disruption and could reach region of resonant conversion intact 
\cite{Prabhu2020, Buckley2021}. So it is still possible that some sub-population of FRBs could be explained by  this process.


\section{Axions and cooling of white dwarfs}

Already in 1986 it was proposed that axions can be emitted by WDs due to bremsstrahlung process when electrons scatter off nucleons (or nuclei) \cite{1986PhLB..166..402R}. This additional energy loss might contribute to WD cooling and so can be detected (directly or indirectly) by different methods. Since publication of the original idea (when some constraints on the axion parameters have been made already) many researchers advanced this approach in many ways. In this section we present and summarize main recent results in this field of astroparticle physics trying to mention different methods of constraining axion parameters via astronomical observations involving WDs. 

The first approach is quite similar to the one described in Sec. 3 above for the case of NSs. 
Axions emitted in hot and dense WD interiors can be converted to photons in strong magnetic fields of MWDs
\cite{2019PhRvL.123f1104D} (see therein references to earlier similar ideas in application to other types of astronomical sources).   
In MWDs surface field can reach $10^9$~G \cite{2020AdSpR..66.1025F}. In such strong field axions can turn to X-ray photons via already discussed above 
Primakoff effect.
\footnote{An opposite process --- conversion of thermal surface photons to axions in magnetospheres of MWDs, --- was used by \cite{2011PhRvD..84h5001G} to put stringent limits on $g_{a\gamma \gamma}$ by observation of linear polarization of optical emission from MWDs.}  Production of X-ray emission is related to the fact that temperatures at WDs interiors can be of the order of $10^7$~K. In comparison with NSs, WDs have one important advantage: their surface temperatures are too low to produce significant flux of thermal X-ray photons of keV energy. Thus, all detected photons can be related to conversion of axions. 

Expected luminosity is non-negligible, in \cite{2019PhRvL.123f1104D} the authors provide the following estimate:

\begin{equation}
L_{ax}= 1.6\times 10^{-4} \left(\frac{g_{aee}}{10^{-13}}\right)^2 \left(\frac{M}{1\, M_\odot}\right) \left(\frac{T_\mathrm{c}}{10^7 \, \mathrm{K}}\right)^4\, L_\odot.
\end{equation}
Here $M$ is WD mass, and $T_\mathrm{c}$ --- its central temperature. 
If all axions are converted to X-ray photons in the WD's magnetosphere, this would correspond to flux $\approx 5 \times 10^{-13}\, d_{100}^{-2}$~erg~cm$^{-2}$~s$^{-1}$, where $d_{100}$ is distance normalized to 100 pc (we neglect interstellar absorption which is reasonable for near-by sources in keV range). Such fluxes are well within reach by modern X-ray observatories like {\it Chandra} or {\it XMM-Newton}, and in future {\it Athena} will bring much more examples. Of course, conversion efficiency is far from ideal. Still, X-ray data on MWD RE J0317-853 obtained with {\it Suzaku} satellite allowed the authors of \cite{2019PhRvL.123f1104D}  to obtain a strong limit on low-mass axions parameters. Later, dedicated observations of the same source produced even better constraint.

The idea proposed in \cite{2019PhRvL.123f1104D}  was recently realized in the study by Dessert et al. \cite{2021arXiv210412772D}. These authors analysed {\it Chandra} observations of MWD RE J0317-853 at $\sim30$~pc from the Sun. No X-ray emission was detected. Accurate modeling and usage of recent data allowed to obtain a better estimate of the central temperature: $\approx 1.4$~keV. Altogether this allowed to place an upper limit $ g_{a\gamma \gamma}\lesssim 5.5\times 10^{-13}\sqrt{C_{a\gamma \gamma}/C_{aee}}$~GeV$^{-1}$. For low-mass axions ($m_a\lesssim 10^{-5}$~eV) this constraint is much better than the one obtained by CAST \cite{2017NatPh..13..584A}. Future observations with new generation of space X-ray observatories might allow to increase this limit substantially, or even to detect the effect.

If axions are important for WD cooling, then this effect influences not only individual sources, but also general properties of the whole population of these compact objects. In particular, luminosity distribution of WDs --- so-called, luminosity function, $n(L)$ \cite{2018MNRAS.478.2569I}:

\begin{equation}
    n(L)=\int^{M_{up}}_{M_{low}} \Phi(M)\Psi(\Delta t)\tau_{cool}(L,M) dM.
    \label{lumfunc}
\end{equation}
In this expression $L$ and $M$ are luminosity and mass of a WD, respectively. $\Phi(M)$ is the initial mass function and $\Psi(t)$ --- star formation rate. Characteristic time scale is defined as $\tau_{cool}=dt/dM_{bol}$, where $M_{bol}$ is absolute bolometric magnitude. Finally, $M_{up}$ and $M_{low}$ are stellar masses that define the range of WD formation in a given population. Each stellar mass has one-to-one correspondence with a WD mass. 
 
 In eq.(\ref{lumfunc}) time interval $\Delta t$ is the difference between the age of population under consideration (can be equal to the Galactic age) and sum of the cooling age (time necessary to reach the given luminosity $L$) and lifetime of the progenitor: $\Delta t = t_{pop} - (t_{cool} (L,M) + t_{pro}(M))$. The minimum progenitor mass can be defined from equation: $t_{pop}=t_{cool}(L, M_{low})+t_{pro}(M_{low})$ (stars with lower masses had no time to evolve up to WD formation, yet).
 
 For a given starformation history standard model of stellar evolution allows to calculate birth rate of WDs of different masses along the lifetime of the analysed population. Then, using standard model of WD cooling we can obtain the luminosity function and compare it with observational data. If they coincide, then there is no necessity to introduce additional cooling agents.\footnote{Axion cooling can also modify stellar evolution and, thus, indirectly influence WD luminosity function, see e.g. \cite{2008LNP...741...51R}.} Oppositely, deviations can be attributed, for example,  to the effect of axions.

Number of known WDs and precision of determination of their parameters rapidly grow in our century thanks to many surveys, especially SDSS and Gaia (see, e.g. \cite{2019MNRAS.482.4570G} and references therein). This allows to determine with high confidence different features, including change of the slope of the luminosity function due to neutrino cooling in hot WDs and effects of crystallization in old (and cold) sources. However, some uncertainties prevent equally solid conclusion about the role of axions, as it is more subtle. 

In the first place, the history of WD formation rate and details of their Galactic distribution  are not known well enough. In particular, recent (1-2 Gyrs) bursts of starformation can modify the luminosity function and result in fallacious conclusions about additional cooling agents. This was recently studied in \cite{2018MNRAS.478.2569I}. 

In their previous studies \cite{2008ApJ...682L.109I} these authors already came to the conclusion that inclusion of axion cooling helps to improve coincidence between observational data and theoretical luminosity functions for absolute bolometric magnitude $M_{bol}\sim 6-12$. The same range of $M_{bol}$, where models without axion cooling slightly overpredict the number of sources, was studied also in \cite{2018MNRAS.478.2569I}. The new analysis accounting for possible episodes of intensive starformation confirms the older result favouring axions with masses about few meV and (electron) coupling constant $\sim$few$\times 10^{-13}$~GeV$^{-1}$. Similar behaviour of luminosity functions for WDs in the Galactic halo, thin and thick disks provides support for the authors' conclusions, as starformation histories of these three elements of the Galactic structure might not be strongly correlated.
Still, uncertainties remain, and new observations are necessary to improve data on the WD luminosity function and new studies of stellar and WD properties would be beneficial too. 

As we already mentioned above, axion cooling influences stellar evolution: stars contract due to additional channel of energy losses from the region of nuclear reactions. Thus, temperature is increased as well as the rate of reactions. The star evolves faster and its helium core grows more massive. On other hand, at the asymptotic giant branch increase of the helium core mass is smaller in the presence of massive axion-like particles. All this results on the mass of a WD, i.e., the relation between initial stellar mass and the compact object mass (IFMR) is modified. 

Properties of the IFMR were studied recently in \cite{Dolan2021}.
These authors slightly improve constraints in comparison to those obtained from helium burning  (so-called horizontal branch) stars. These objects have dense hot cores and their lifetime can be estimated from observations. Additional cooling due to axions can significantly modify duration of helium burning in contradiction with observations. This excludes massive axion-like particles in the range $\sim1 - 100$~keV with axion-photon coupling constant $g_{a\gamma \gamma} \gtrsim 10^{-10} $~GeV$^{-1}$. 
In \cite{Dolan2021} the authors enlarge the forbidden area in the region of so-called cosmological triangle with axion masses about hundreds of keV and coupling constant $\sim 10^{-5}$~GeV$^{-1}$ analysing helium core growth at the asymptotic giant branch stage. 

In helium burning shells of asymptotic giant branch stars temperature is about 15-20 keV. 
At these conditions Primakoff process and photon-photon interaction can result in production of axion-like particles  even if their masses  $\gtrsim$ than few tens of keV.
Additional cooling leads to faster evolution and more active mixing of matter in outer layers. The latter process results in smaller amount of helium available for burning in the shell and so, the mass of carbon-oxygen core grows slower. Finally, the WD originated from this core is lighter than the one which could be formed in the standard scenario without additional cooling. 

In \cite{Dolan2021} the authors used the IFMR based on 14 double WD wide binary systems. This IFMR covers the range progenitor masses from 2 up to 7$M_\odot$. If massive ($\sim10$--~few$\times 100$~keV) axion-like particles are considered for $g_{a\gamma \gamma}\gtrsim 10^{-10}$~Gev$^{-1}$ theoretical IFMR starts to be out of limits defined by observations in the region of progenitor masses $\sim4-7\, M_\odot$ (excluding very large values $g_{a\gamma \gamma} \gtrsim 10^{-3}$GeV$^{-1}$ where other effects become important as axion-like particles do not leave the helium burning shell, but decay inside it). Accounting for rotation slightly softens the constraint, but still it increases forbidden area in comparison with limits based on analysis of core helium-burning stars. 

May be the most interesting constraint using WDs is obtained from the analysis of a pulsating source belonging to the so-called ZZ Ceti class (on formation of this type of WDs, see e.g. \cite{2002MNRAS.330..685A} and references therein). These WDs are situated in the instability strip of the Hertzsprung-Russel diagram. They are mid-aged ($<10^9$~yrs) and so still warm ($>10^4$~K) objects with convective envelopes containing zones of partial ionization and hydrogen atmospheres. 

The constraint is based on observations of pulsation period evolution of G117–B15A --- a near-by ($\approx 57.5 $~pc) WD with temperature $T\approx 12,000$~K and mass $\sim 0.5\, M_\odot$. This source is monitored since mid-70s. ZZ Ceti stars increase pulsation period as they cool down. In the simplest model $\dot P/P \propto  - \dot T/T$ (e.g. \cite{1992ApJ...392L..23I}).
G117–B15A has the period equal to 215.2 seconds and it is increasing with the rate $\dot P=5.47\pm 0.82 \times 10^{-15}$~s~s$^{-1}$ \cite{2021ApJ...906....7K}, less than 10\% of thus value can be attributed to the proper motion of the source.
Already in \cite{1992ApJ...392L..23I} it was proposed that large value of the period derivative ($\dot P=12\pm 3.5 \times 10^{-15}$~s~s$^{-1}$ at that time) can be explained by additional cooling due to axions. In that paper the authors obtained the value of axion mass about a few meV.

In the new study  \cite{2021ApJ...906....7K} the authors estimate axion mass as $m_a \cos^2 \beta =19.9^{+2.1}_{-3.1}$~meV, where $\cos^2\beta$ is a free parameter\footnote{This model-dependent parameter links electron coupling constant $g_{ae}$ to axion mass $m_a$ (in units of eV) in the DFSZ model: $g_{ae}=2.83\times10^{-11}m_a/\cos^2\beta$ and  is usually set equal to unity.}. This value might be considered as an upper limit as several uncertainties remain. 



\section{Conclusions}

Perspectives for laboratory detection of DM particles and axions in near future are not very bright. Thus, most probably astronomical data will not only have a chance to continue producing the best upper limits, but also are number one candidate to provide direct evidence in favour of existence of such particles. Various approaches related to different types of sources are used in astroparticle physics. But continuously NSs and WDs are among best sites to look for effects linked to new physics. 

In the review we presented several lines of considerations to identify observational tests of existence of exotic particles predicted by theories. We expect that in near future some of them have chances to provide positive results. To name a few:
\begin{itemize}
\item deep  surveys can identify dim old  sources related to isolated NSs heated by DM particles annihilation; 
\item dedicated searches for radio emission related to Primakoff processes can result in its detection; \item finally, detailed studies (including polarization) of surface X-ray and optical emission of cooling NSs also can demonstrates effects related to the existence of axions.  
\end{itemize}

Advances in technology promise persistent progress in sensitivity of astronomical observations in next decades. All this raises the chances to discover new particles by astronomical means

\authorcontributions{All authors contributed equally to secs. 1 and 5. Sec. 2 was mainly written by PT, sec. 3 --- by MP, and sec. 4 --- by SP.
All authors have read and agreed to the published version of the manuscript. }

\funding{MP and SP were supported by the Ministry of Science and Higher Education of Russian Federation under the contract 075-15-2020-778 in the framework of the Large Scientific Projects program within the national project ``Science''.}

\acknowledgments{The authors are grateful to Sergey Troitsky and Alexey Zhuravlev for helpful discussions.}

\conflictsofinterest{The authors declare no conflict of interest.} 



\abbreviations{The following abbreviations are used in this manuscript:\\

\noindent 
\begin{tabular}{@{}ll}
ALP & Axion-like particle\\
BH & black hole\\
BEC & Bose-Einstein condensate\\
CP & Charge parity\\
DFSZ & Dine-Fischler-Srednicki-Zhitnitsky model\\
DM  & Dark matter \\
EM & Electro-magnetic\\
FRB & Fast radio burst\\
IFMR & Initial to final mass ratio \\
KSVZ & Kim-Shifman-Vainshtein-Zakharov model\\
MWD & Magnetic white dwarf\\
NS & Neutron star\\
NFW & Navarro, Frenk and White model\\
 QCD & Quantum chromodynamics \\
 QED & Quantum electrodymanics\\
SM & Standard model \\
SMBH & Supermassive black hole\\
 WD & White dwarf\\
 WIMP & Weakly interacting massive particle \\
 XDINS & X-ray dim isolated neutron stars\\
 
\end{tabular}}

\appendixtitles{no} 




\reftitle{References}


\externalbibliography{yes}
\bibliography{references}

\end{paracol}
\end{document}